\documentstyle[seceq]{ptptex}

\notypesetlogo  

\title{
Triplet-Doublet Splitting, Proton Stability and an Extra Dimension
}

\author{
Yoshiharu {\sc Kawamura}\footnote{E-mail:
haru@azusa.shinshu-u.ac.jp}
}

\inst{
Department of Physics, Shinshu University, Matsumoto 390-8621, Japan
}

\recdate{
December 11, 2000
}

\abst{
We propose a new possibility to reconcile the coupling unification
scenario with the triplet-doublet mass splitting
based on a 5-dimensional supersymmetric model with $SU(5)$
gauge symmetry.
It is shown that the minimal supersymmetric standard model
is derived on a 4-dimensional wall 
through compactification on $S^1/(Z_2 \times Z_2')$.
}

\begin{document}

\maketitle

\section{Introduction}

The minimal supersymmetric standard model (MSSM) is 
the most promising model to describe physics beyond
the Standard Model (SM).
An advantageous feature of the MSSM is that the 
gauge coupling constants meet at $M_X = 2.1 \times 10^{16}$GeV 
if the superpartners and Higgs particles
exist below or around $O(1)$ TeV.\cite{unification}
This fact leads to the possibility that gauge interactions in the MSSM are
unified under a
simple gauge group, such as $SU(5)$.
If this is indeed the case, then the theory can be described as a
supersymmetric grand unified theory (SUSY GUT).\cite{SUSY GUT}
This scenario is very attractive, but, in general, it suffers from
problems related to Higgs multiplets.
For example, in the minimal SUSY $SU(5)$ GUT, a fine tuning is required 
to obtain the $SU(2)_L$ doublet Higgs multiplets with the weak scale mass
such that the colored Higgs multiplets remain sufficiently heavy to suppress
dangerous nucleon decay (the triplet-doublet splitting problem). 
There have been several interesting proposals to solve this problem
through the extension of the model.\cite{TD1,TD2,TD3,TD4,TD5,TD6}

In this paper, we propose a new possibility to reconcile the coupling
unification
scenario with the triplet-doublet mass splitting.
The coupling unification originates from the existence of a unified gauge
symmetry, $G$,
in a high-energy theory on a higher-dimensional space-time.
The full symmetry is not realized in the low-energy physics, where light
particles
play an essential role.
The symmetry $G$ is reduced by the presence of non-universal values of the
intrinsic parity 
on a compact space among components in each multiplet of $G$.
We show that the SUSY part of the Lagrangian in the MSSM is derived 
on a 4-dimensional (4D) wall through compactification upon the orbifold
$S^1/(Z_2 \times Z_2')$,  with a suitable assignment
of $Z_2 \times Z_2'$ parity, from a 5D SUSY model based on $G=SU(5)$.
No particles appear other than the MSSM particles in the massless state,
and the triplet-doublet mass splitting is realized 
by the $Z_2 \times Z_2'$ projection.
We also discuss proton stability in our model.

This paper is organized as follows. 
In the next section, we explain the construction of the orbifold 
$S^1/(Z_2 \times Z_2')$ and describe an intrinsic parity on this compact space.
Starting from 5D SUSY $SU(5)$ GUT with minimal particle content,
we derive the SUSY part of the Lagrangian in the 4D theory and
discuss the mass spectrum and its phenomenological implications in $\S$3. 
Section 4 is devoted to conclusions and discussion.

\section{${\it S^1/(Z_2 \times Z_2')}$ and parity}

The space-time is assumed to be factorized into a product of 
4D Minkowski space-time $M^4$ and the orbifold $S^1/(Z_2 \times
Z_2')$,\footnote{
Recently, Barbieri, Hall and Nomura have
constructed a constrained standard model 
upon a compactification of a 5D SUSY model on 
the orbifold $S^1/(Z_2 \times Z_2')$.\cite{BHN}
They used $Z_2 \times Z_2'$ parity to reduce SUSY.
There are also several works on model building 
through a reduction of SUSY \cite{AMQ,HW,MP,PQ} and a gauge symmetry \cite{K} 
by the use of $Z_2$ parity.}
whose coordinates are denoted
by $x^\mu$ ($\mu = 0,1,2,3$) and $y$ $(=x^5)$, respectively.
The 5D notation $x^M$ ($M = 0,1,2,3,5$) is also used.
We first construct the orbifold $S^1/Z_2$ by dividing a circle $S^1$
of radius $R$ with a $Z_2$ transformation which acts on $S^1$
according to $y \to -y$.
This compact space is regarded as the interval $[-\pi R, 0]$ with a length
of $\pi R$.
The orbifold $S^1/(Z_2 \times Z_2')$ is obtained by dividing $S^1/Z_2$
with another $Z_2$ transformation, denoted by $Z_2'$, which acts on $S^1/Z_2$
according to $y' \to -y'$, where $y' \equiv y + {\pi R \over 2}$.
This compact space is regarded as 
the interval $[0, {\pi R \over 2}]$ with a length of 
${\pi R \over 2}$.
There are two 4D walls placed at
the fixed points $y'=0$ and $y'={\pi R \over 2}$ 
(or $y=-{\pi R \over 2}$ and $y=0$) on $S^1/(Z_2 \times Z_2')$.

The intrinsic $Z_2 \times Z_2'$ parity of the 5D bulk field $\phi(x^\mu, y)$
is defined by the transformation
\begin{eqnarray}
\phi(x^\mu, y) \to \phi(x^\mu, -y) &=& P \phi(x^\mu, y) ,
\label{P-tr}\\
\phi(x^\mu, y') \to \phi(x^\mu, -y') &=& P' \phi(x^\mu, y') .
\label{P'-tr}
\end{eqnarray}
The Lagrangian should be invariant under the $Z_2 \times Z_2'$ transformation.
By definition, $P$ and $P'$ possess only the eigenvalues 1 and $-1$.
We denote the fields that are simultaneous eigenfunctions of
these operators as $\phi_{++}$, $\phi_{+-}$, $\phi_{-+}$
and $\phi_{--}$, where the first subscript corresponds to the
eigenvalue of $P$ and the second to $P'$.
The fields $\phi_{++}$, $\phi_{+-}$, $\phi_{-+}$and $\phi_{--}$ are Fourier
expanded as
\begin{eqnarray}
  \phi_{++} (x^\mu, y) &=& 
       \sqrt{2 \over {\pi R}} 
      \sum_{n=0}^{\infty} \phi^{(2n)}_{++}(x^\mu) \cos{2ny \over R} ,
\label{phi++exp}\\
  \phi_{+-} (x^\mu, y) &=& 
       \sqrt{2 \over {\pi R}} 
      \sum_{n=0}^{\infty} \phi^{(2n+1)}_{+-}(x^\mu) \cos{(2n+1)y \over R} ,
\label{phi+-exp}\\
  \phi_{-+} (x^\mu, y) &=& 
       \sqrt{2 \over {\pi R}}
      \sum_{n=0}^{\infty} \phi^{(2n+1)}_{-+}(x^\mu) \sin{(2n+1)y \over R}  ,
\label{phi-+exp}\\
  \phi_{--} (x^\mu, y) &=& 
       \sqrt{2 \over {\pi R}}
      \sum_{n=0}^{\infty} \phi^{(2n+2)}_{--}(x^\mu) \sin{(2n+2)y \over R}  ,
\label{phi--exp}
\end{eqnarray}
where $n$ is an integer, and each field $\phi^{(2n)}_{++}(x^\mu)$,
$\phi^{(2n+1)}_{+-}(x^\mu)$, $\phi^{(2n+1)}_{-+}(x^\mu)$
and $\phi^{(2n+2)}_{--}(x^\mu)$
acquire a mass ${2n \over R}$, 
${2n+1 \over R}$, ${2n+1 \over R}$ and 
${2n+2 \over R}$ upon compactification. 
Note that 4D massless fields appear only in $\phi_{++}(x^\mu, y)$.
We find that some fields vanish on the wall, for example,
$\phi_{+-}(x^\mu, -{\pi R \over 2})$ 
$= \phi_{--}(x^\mu, -{\pi R \over 2})=0$
on the wall placed at $y = -{\pi R \over 2}$,
and $\phi_{-+}(x^\mu, 0)$ $= \phi_{--}(x^\mu, 0)=0$
on the other wall placed at $\displaystyle{y = 0}$.

Let us study the case in which a field $\Phi(x^\mu, y)$ is an $N$-plet
under some symmetry group $G$.
The components of $\Phi$ are denoted by $\phi_k$ as
$\Phi = (\phi_1, \phi_2, ..., \phi_N)^T$.
The $Z_2$ transformation of $\Phi$ takes the same form as 
(\ref{P-tr}), but in this case $P$ is an $N \times N$ matrix
\footnote{$P$ is a unitary and hermitian matrix.} 
that satisfies $P^2 = I$, where $I$ is the unit matrix.
The $Z_2$ invariance of the Lagrangian
does not necessarily require that $P$ be $I$ or $-I$.
Unless all components of $\Phi$ have a common $Z_2$ parity (i.e.,
if $P \neq \pm I$), a symmetry reduction occurs upon compactification,
because of the lack of zero modes in components with odd parity.\cite{K}
The same property holds in the case with $Z_2'$ parity.

\section{A model with {\it SU(5)} gauge symmetry}

We now study 5D SUSY $SU(5)$ GUT with minimal particle content
and non-universal parity assignment.
We assume that the vector supermultiplet $V$ and two kinds of
hypermultiplets $H^s$ ($s=1,2$) exist 
in the bulk $M^4 \times S^1/(Z_2 \times Z_2')$.
The vector multiplet $V$ consists of a vector boson $A_M$,
two bispinors $\lambda^i_L$ ($i = 1,2$), and a real scalar $\Sigma$,
which together form an adjoint representation ${\bf 24}$ of $SU(5)$, and
the hypermultiplets $H^s$ consist of 
two complex scalar fields and two Dirac fermions 
$\psi^s = (\psi_L^s, \psi_R^s)^T$, which are equivalent to
four sets of chiral supermultiplets: 
$H^1 =$ $\{H_{\bf 5} \equiv (H_1^1, \psi_L^1)$,
$\hat{H}_{\bar{{\bf 5}}} \equiv (H_2^1, \bar{\psi}_R^1)\}$ and
$H^2 =$ $\{\hat{H}_{\bf 5} \equiv (H_1^2, \psi_L^2)$,
$H_{\bar{{\bf 5}}} \equiv (H_2^2, \bar{\psi}_R^2)\}$.
The hypermultiplets $H_{\bf 5}$ and $\hat{H}_{\bf 5}$ 
($\hat{H}_{\bar{{\bf 5}}}$ and $H_{\bar{{\bf 5}}}$)
form a fundamental representation ${\bf 5}$ ($\bar{\bf 5}$).
We assume that our visible world is a 4D wall
fixed at $y=0$ (We refer to as $\lq\lq$wall I")
and that three families of quark and lepton chiral supermultiplets,
$3\{\Phi_{\bar{\bf 5}} + \Phi_{\bf 10}\}$, are located on this wall.
(Here and hereafter the family index does not appear.)
That is, matter fields contain no excited states along the
$S^1/(Z_2 \times Z_2')$ direction.

The gauge invariant action is given by
\begin{eqnarray}
 S &=& \int  {\cal L}^{(5)} d^5x  + {1 \over 2} \int \delta(y) {\cal
L}^{(4)} d^5x 
\nonumber \\
 &~& + (\mbox{terms from a brane fixed at } y = -\pi R) , \\
 {\cal L}^{(5)} &=& {\cal L}^{(5)}_{YM} + {\cal L}^{(5)}_{H} ,\\
 {\cal L}^{(5)}_{\rm YM} &=& -{1 \over 2} {\rm Tr} F_{MN}^2
  + {\rm Tr} |D_M \Sigma|^2 
  + {\rm Tr} (i\bar{\lambda}_i \gamma^M D_M \lambda^i)
  - {\rm Tr} (\bar{\lambda}_i [\Sigma, \lambda^i])  ,\\
 {\cal L}^{(5)}_{\rm H} &=& |D_M H_i^s|^2 
    + i\bar{\psi}_s \gamma^M D_M \psi^s
  - (i \sqrt{2} g_{(5)} \bar{\psi}_s \lambda^i H_i^s + \mbox{h.c.}) 
\nonumber \\
&~& - \bar{\psi}_s \Sigma \psi^s - H^{\dagger i}_s \Sigma^2 H_i^s 
- {g_{(5)}^2 \over 2} \sum_{m, A} 
\left(H_s^{\dagger i} (\sigma^m)_i^j T^A H_j^s \right)^2  ,\\
{\cal L}^{(4)} &\equiv&  \sum_{3 {\rm families}} 
\int d^2\bar{\theta} d^2\theta (\Phi^\dagger_{\bar{\bf 5}} e^{2 g_{(5)} V^A
T^A} \Phi_{\bf \bar{5}} 
+ \Phi^\dagger_{\bf 10} e^{2 g_{(5)} V^A T^A} \Phi_{\bf 10}) 
\nonumber \\
&~& + \sum_{3 {\rm families}} \int d^2 \theta (f_{U(5)} H_{\bf 5} \Phi_{\bf
10} \Phi_{\bf 10} 
+ \hat{f}_{U(5)} \hat{H}_{\bf 5} \Phi_{\bf 10} \Phi_{\bf 10} 
\nonumber \\
&~& ~~~~~~~~~ + f_{D(5)} H_{\bar{\bf 5}} \Phi_{\bf 10} \Phi_{\bar{\bf 5}} 
+ \hat{f}_{D(5)} \hat{H}_{\bar{\bf 5}} \Phi_{\bf 10} \Phi_{\bar{\bf 5}}) +
\mbox{h.c.} ,
\end{eqnarray}
where $\lambda^i \equiv (\lambda_L^i, \epsilon^{ij} \bar{\lambda}_{Lj})^T$,
$D_M \equiv \partial_M - i g_{(5)} A_M(x^\mu, y)$,
$g_{(5)}$ is a 5D gauge coupling constant, 
the $\sigma^m$ are Pauli matrices, the $T^A$ are $SU(5)$ gauge generators,
$V^A T^A$ is an $SU(5)$ vector supermultiplet,
and $f_{U(5)}$, $f_{D(5)}$, $\hat{f}_{U(5)}$ and $\hat{f}_{D(5)}$
are 5D Yukawa coupling matrices.
If we impose $Z_2$ invariance\footnote{
This $Z_2$ is a discrete subgroup of $SU(2)_H$,
which is one of the global symmetries of ${\cal L}^{(5)}$.
The Higgs bosons transform as ${\bf 2}$ under $SU(2)_H$.}
under $H_{\bf 5} \leftrightarrow \hat{H}_{\bf 5}$ and 
$H_{\bar{\bf 5}} \leftrightarrow \hat{H}_{\bar{\bf 5}}$
on ${\cal L}^{(4)}$, 
the relations $f_{U(5)} = \hat{f}_{U(5)}$ and $f_{D(5)} = \hat{f}_{D(5)}$
are derived.
The representations of $\Phi_{\bar{\bf 5}}$ and $\Phi_{\bf10}$ 
are $\bar{\bf 5}$ and ${\bf 10}$, respectively.
In ${\cal L}^{(4)}$, the bulk fields are replaced
by fields including the Nambu-Goldstone boson $\phi(x^\mu)$ at the wall I, 
$V^A(x^\mu, \theta, \bar{\theta}, y=\phi(x^\mu))$ 
and $H^s(x^\mu, \theta, y=\phi(x^\mu))$.\cite{BKNY}
In the above action, we assume that there is a symmetry such as $R$ parity
to forbid the term $\Phi_{\bar{\bf 5}} \Phi_{\bar{\bf 5}} \Phi_{\bf 10}$
from appearing in the superpotential, which induces rapid proton decay.
The Lagrangian is invariant under
the $Z_2$ transformation
\begin{eqnarray}
 &~& A_{\mu}(x^\mu, y) \to A_{\mu}(x^\mu, -y) = 
P A_{\mu}(x^\mu, y) P^{-1} , \nonumber \\
 &~& A_{5}(x^\mu, y) \to A_{5}(x^\mu, -y) = 
- P A_{5}(x^\mu, y) P^{-1} , \nonumber \\
 &~& \lambda_L^1(x^\mu, y) \to \lambda_L^1(x^\mu, -y) = 
- P \lambda_L^1(x^\mu, y) P^{-1} , \nonumber \\
 &~& \lambda_L^2(x^\mu, y) \to \lambda_L^2(x^\mu, -y) = 
P \lambda_L^2(x^\mu, y) P^{-1} , \nonumber \\
 &~& \Sigma(x^\mu, y) \to \Sigma(x^\mu, -y) = 
- P \Sigma(x^\mu, y) P^{-1} , \nonumber \\
 &~& H_{\bf 5}(x^\mu, y) \to H_{\bf 5}(x^\mu, -y) 
 = P H_{\bf 5}(x^\mu, y) , \nonumber \\
 &~& \hat{H}_{\bar{\bf 5}}(x^\mu, y) \to \hat{H}_{\bar{\bf 5}}(x^\mu, -y) 
 = - P \hat{H}_{\bar{\bf 5}}(x^\mu, y) , \nonumber \\
 &~& \hat{H}_{\bf 5}(x^\mu, y) \to \hat{H}_{\bf 5}(x^\mu, -y) 
 = - P \hat{H}_{\bf 5}(x^\mu, y) , \nonumber \\
 &~& H_{\bar{\bf 5}}(x^\mu, y) \to H_{\bar{\bf 5}}(x^\mu, -y) 
 = P H_{\bar{\bf 5}}(x^\mu, y) 
\label{P-tr2}
\end{eqnarray}
and under the $Z_2'$ transformation, obtained by replacing $y$ and $P$ by
$y'$ and $P'$ in the above.

When we use $P={\rm diag}(1,1,1,1,1)$ and $P'={\rm
diag}(-1,-1,-1,1,1)$,\footnote{
The exchange of $P$ and $P'$ is equivalent to 
the exchange of two walls.
The origin of this specific $Z_2 \times Z_2'$ parity assignment is unknown, and
we believe that it will be explained in terms of
some yet to be constructed underlying theory.}
the $SU(5)$ gauge symmetry is reduced to that of
the Standard Model, $G_{\rm SM} \equiv SU(3) \times SU(2)
\times U(1)$, in the 4D theory.\footnote{
Our symmetry reduction mechanism is 
different from the Hosotani mechanism.\cite{H}
In fact, the Hosotani mechanism does not work in our case,
because $A_{5}^{a}(x^\mu,y)$ has odd parity, as seen from (\ref{P-tr2}),
and its VEV should vanish.}
This is because the boundary conditions on $S^1/(Z_2 \times Z_2')$ 
do not respect $SU(5)$ symmetry,
as we see from the relations for the gauge generators $T^A$ 
$(A = 1, 2, \cdots,24)$,
\begin{eqnarray}
  P' T^a P'^{-1} = T^a , ~~
  P' T^{\hat{a}} P'^{-1} = -T^{\hat{a}}  .
\end{eqnarray}
The $T^a$ are gauge generators of $G_{\rm SM}$,
and the $T^{\hat{a}}$ are the other gauge generators.
The parity assignment and mass spectrum after compactification are given in
Table I.
Each Higgs multiplet is divided into two pieces: 
$H_{\bf 5}$ ($\hat{H}_{\bar{\bf 5}}$, $\hat{H}_{\bf 5}$, $H_{\bar{\bf 5}}$) is
divided into the colored triplet piece, $H_C$ 
($\hat{H}_{\bar{C}}$, $\hat{H}_{C}$, $H_{\bar{C}}$),
and the $SU(2)$ doublet piece, $H_u$ ($\hat{H}_{d}$, $\hat{H}_{u}$, $H_{d}$).
In the second column, we give the $SU(3) \times SU(2)$ quantum numbers
of the 4D fields.
In the third column, $(\pm, \pm)$ and $(\pm, \mp)$ 
denote the eigenvalues $(\pm 1, \pm1)$ and $(\pm 1, \mp1)$ 
of $Z_2 \times Z_2'$ parity, respectively.
In the fourth column, $n$ represents 0 or a positive integer.
The massless fields are $(A_{\mu}^{a(0)}$, $\lambda^{2a(0)})$ and 
$(H_u^{(0)}$, ${H}_d^{(0)})$, which are equivalent to 
the gauge multiplets and the weak $SU(2)$ doublet Higgs multiplets 
in the MSSM, respectively.
We find that the triplet-doublet mass splitting of the Higgs multiplets is
realized
by projecting out zero modes of the colored components.
\begin{table}[t]
\caption{Parity and mass spectrum at the tree level.}
\renewcommand{\arraystretch}{1.6}
\begin{center}
\begin{tabular}{l|l|l|l}
\hline\hline
4D fields & Quantum numbers & $Z_2 \times Z_2'$ parity & Mass \\
\hline
$A_{\mu}^{a(2n)} $, $\lambda^{2a(2n)}$ & $({\bf 8}, {\bf 1}) + ({\bf 1},
{\bf 3})
 + ({\bf 1}, {\bf 1})$ & $(+, +)$ & $\displaystyle{{2n \over R}}$ \\
$A_{\mu}^{\hat{a}(2n+1)}$, $\lambda^{2\hat{a}(2n+1)}$ 
&  $({\bf 3}, {\bf 2}) + (\bar{\bf 3}, {\bf 2})$
& $(+, -)$ & $\displaystyle{{2n+1 \over R}}$ \\
\hline
$A_{5}^{a(2n+2)}$, $\Sigma^{a(2n+2)}$, $\lambda^{1a(2n+2)}$ 
& $({\bf 8}, {\bf 1}) + ({\bf 1}, {\bf 3})
 + ({\bf 1}, {\bf 1})$ & $(-, -)$ & $\displaystyle{{2n+2 \over R}}$ \\
$A_{5}^{\hat{a}(2n+1)}$, $\Sigma^{\hat{a}(2n+1)}$, 
$\lambda^{1\hat{a}(2n+1)}$ &  
$({\bf 3}, {\bf 2}) + (\bar{\bf 3}, {\bf 2})$
& $(-, +)$ & $\displaystyle{{2n+1 \over R}}$ \\
\hline
$H_C^{(2n+1)}$ & $({\bf 3}, {\bf 1})$ & $(+, -)$ & $\displaystyle{{2n+1
\over R}}$ \\
$H_u^{(2n)}$ & $({\bf 1}, {\bf 2})$ & $(+, +)$ & $\displaystyle{{2n \over
R}}$ \\
$\hat{H}_{\bar{C}}^{(2n+1)}$ & $(\bar{{\bf 3}}, {\bf 1})$ & $(-, +)$ 
& $\displaystyle{{2n+1 \over R}}$ \\
$\hat{H}_d^{(2n+2)}$ & $({\bf 1}, {\bf 2})$ & $(-, -)$ &
$\displaystyle{{2n+2 \over R}}$ \\
\hline
$\hat{H}_C^{(2n+1)}$ & $({\bf 3}, {\bf 1})$ & $(-, +)$ &
$\displaystyle{{2n+1 \over R}}$ \\
$\hat{H}_u^{(2n)}$ & $({\bf 1}, {\bf 2})$ & $(-, -)$ & $\displaystyle{{2n+2
\over R}}$ \\
$H_{\bar{C}}^{(2n+1)}$ & $(\bar{{\bf 3}}, {\bf 1})$ & $(+, -)$ 
& $\displaystyle{{2n+1 \over R}}$ \\
${H}_d^{(2n)}$ & $({\bf 1}, {\bf 2})$ & $(+, +)$ & $\displaystyle{{2n \over
R}}$ \\
\hline
\end{tabular}
\end{center}
\end{table}

After integrating out the fifth dimension,
we obtain the 4D Lagrangian density,
\begin{eqnarray}
{\cal L}^{(4)}_{\rm eff} &=& {\cal L}^{(4)}_B + {\cal L}^{(4)} ,\\
{\cal L}^{(4)}_B &\equiv& - {1 \over 4} \int d^2 \theta W^{a(0)} W^{a(0)} +
\mbox{h.c.}
+ \int d^2\bar{\theta} d^2\theta (H^{\dagger(0)}_{u} e^{2 g_U V^{a(0)} T^a}
H_{u}^{(0)} 
\nonumber \\
&~& + H^{\dagger(0)}_{d} e^{2 g_{U} V^{a(0)} T^a} H_{d}^{(0)}) 
 + \cdots , \\
{\cal L}^{(4)} &\equiv& \sum_{3 {\rm families}} 
\int d^2\bar{\theta} d^2\theta  \Phi^\dagger e^{2 g_U V^{a(0)} T^a} \Phi 
\nonumber \\
&~& + \sum_{3 {\rm families}} \int d^2 \theta (f_U H_u^{(0)} Q U^c  
+ f_D H_d^{(0)} Q D^c 
\nonumber \\
&~& ~~~~~~~~ + f_D H_d^{(0)} L E^c)+ \mbox{h.c.} + \cdots ,
\label{4D-L}
\end{eqnarray}
where $V^{a(0)}$ is a vector multiplet of the MSSM gauge bosons
and gauginos, and
the dots represent terms including 
Kaluza-Klein modes.
In this equation, $g_U$ 
$\left(\equiv \sqrt{2 \over \pi R} g_{(5)}\right)$ is a 4D 
gauge coupling constant, $f_U$ 
$\left(\equiv \sqrt{2 \over \pi R} f_{U(5)}\right)$
and $f_D$ 
$\left(\equiv \sqrt{2 \over \pi R} f_{D(5)}\right)$
are 4D Yukawa coupling matrices, 
$Q$, $U^c$ and $D^c$ are quark chiral supermultiplets, and
$L$ and $E^c$ are lepton chiral supermultiplets.
We denote these matter multiplets as $\Phi$ generically in the kinetic term.
With our assignment of $Z_2 \times Z_2'$ parity,
the SUSY part of the Lagrangian density in the MSSM is obtained, with the
exception
of the $\mu$ term.

The theory predicts that coupling constants of $G_{\rm SM}$ are unified 
around the compactification scale $M_C (\equiv 1/R)$
to zero-th order approximation,
as in the ordinary $SU(5)$ GUT:\cite{GUT}
\begin{eqnarray}
&~& g_3 = g_2 = g_1 = g_U , ~~ f_d = f_e = f_D~,
\end{eqnarray}
where $f_d$ and $f_e$ are Yukawa coupling matrices on
down-type quarks and electron-type leptons, respectively. From the precise
measurements at LEP, \cite{LEP} it is natural to identify $M_C$ with 
$M_X = 2.1 \times 10^{16}$ GeV.
In this case, the masses of the Kaluza-Klein excitations are quantized
in units of $M_X$.

Finally, we discuss nucleon stability in our model.\cite{pr-largeR}
It is known that there is a significant contribution to proton decay
from the dimension 5 operator in the minimal SUSY $SU(5)$ GUT.\cite{dim5}
Stronger constraints on the colored Higgs mass $M_{H_C}$ and the sfermion
mass $m_{\tilde{f}}$ have been obtained (e.g. $M_{H_C} > 6.5 \times 10^{16}$ GeV
for $m_{\tilde{f}} < 1$ TeV) from analysis including
a Higgsino dressing diagram with right-handed matter fields.\cite{GN}
In our model, we have diagrams similar to those in the minimal SUSY $SU(5)$
GUT, because quark and lepton supermultiplets couple to the Kaluza-Klein modes
of extra vector supermultiplets and the colored Higgs triplets at the tree
level.
Hence the identification $M_C = M_X$ seems to be 
incompatible with the above constraint,
$M_{H_C} > 6.5 \times 10^{16}$ GeV for $m_{\tilde{f}} < 1$ TeV.
However there is a 
natural way to escape this difficulty.
It is pointed out that there is an exponential suppression factor in the
coupling
to the Kaluza-Klein modes resulting from the brane recoil effect.\cite{BKNY}
There is a possibility that proton stability is guaranteed 
if our 4D wall fluctuates pliantly.\footnote{
It is not obvious whether or not the recoil effect exists for a brane at an
orbifold
fixed point.\cite{BN}}

\section{Conclusions and discussion}

We have proposed a new possibility to reconcile the coupling unification
scenario with the triplet-doublet mass splitting.
The coupling unification originates from the existence of a unified gauge
symmetry
$SU(5)$ in 5D space-time.
The full symmetry is not realized in low-energy physics, where light particles
play an essential role; that is, here this symmetry is reduced by 
the existence of non-universal values of the intrinsic parity 
on a compact space among components in each multiplet of $SU(5)$.
We have shown that the SUSY part of the Lagrangian in the MSSM is 
derived, with the exception of the $\mu$ term, on a 4D wall through
compactification upon 
$S^1/(Z_2 \times Z_2')$ from a 5D SUSY model based on $SU(5)$, 
under the assumption 
that our visible world consists of a 4D wall 
and that matter multiplets live on the wall.
In the sector with renormalizable interactions,
the theory predicts the coupling unification
$g_3 = g_2 = g_1 = g_U$ and $f_d = f_e = f_D$, as in the GUT model.
No particles appear other than the MSSM ones in the massless state,
and the triplet-doublet mass splitting is realized 
through the $Z_2 \times Z_2'$ projection.
Although the quark and lepton multiplets couple to the Kaluza-Klein modes
of extra vector multiplets and the colored Higgs triplets 
at the tree level, there is a possibility that proton stability is guaranteed 
by the appearance of a suppression factor in the coupling
to the Kaluza-Klein modes if our 4D wall fluctuates flexibly.
It is not yet known if the above mechanism works for a brane at an orbifold
fixed point.

There are several problems with our model.
Here we list some of them.
The first three problems are peculiar to the MSSM,
and the others are related to the 4D walls.
The first one involves the question of how to avoid rapid proton decay
that results from the term 
$\Phi_{\bar{\bf 5}} \Phi_{\bar {\bf 5}} \Phi_{\bf 10}$ in the superpotential.
We need a symmetry, such as $R$ parity.
The second problem involves the question of
how to generate the $\mu$ term with a suitable magnitude consistent with
the electro-weak symmetry.
The third problem regards the origin of SUSY breaking.
The fourth problem concerns the necessity of 
non-universal $Z_2 \times Z_2'$ parity, i.e.,
whether there is a selection rule that picks out
a specific $Z_2 \times Z_2'$ parity to break $SU(5)$ down to $G_{\rm SM}$
and whether it is compatible 
with brane fluctuations.
The last problem regards how matter fields are localized on the 4D wall.
We expect that these problems can be solved by a yet unknown underlying theory.

In spite of these problems, it is worthwhile to study 
the relation between the symmetry in the SM
and the characteristics of a compact space for the purpose of constructing
a realistic model.\footnote{
Attempts to construct GUT have been made through dimensional reduction 
over coset space.\cite{coset}}

~~\\
{\footnotesize {\bf Note added}: ~While revising this manuscript, 
we found Ref. \citen{AF} by G.~Altarelli and 
F.~Feruglio and Ref. \citen{HN} by L.~J.~Hall and Y.~Nomura
in the hep-archive.
In the  context of our model, proton decay is analyzed in the former paper
and several phenomenological features, including gauge coupling unification,
are studied in the latter paper.}

\end{document}